\documentclass[twocolumn,amsmath,amssymb,showkeys,showpacs,pra,english]{revtex4}
\makeatletter

\usepackage{graphicx}
\usepackage{amsmath}
\usepackage{amssymb}
\usepackage{babel}
\usepackage{color}
\makeatother

\usepackage{graphicx}
\usepackage{dcolumn}
\usepackage{bm}

\begin{document}

\title{Photon recycling in Fabry-Perot micro-cavities based on Si$_3$N$_4$
waveguides.}
\author{F. Riboli}
\email[]{riboli@science.unitn.it}
\author{A. Recati$^a$}
\author{N. Daldosso}
\author{L. Pavesi}
\affiliation{Department of Physics, University of Trento, via
Sommarive 14, Povo, I-38050 Trento, Italy } \affiliation{$^a$ CRS
BEC-INFM, via Sommarive 14, Povo, I-38050 Trento, Italy}

\author{G. Pucker}
\author{A. Lui}
\affiliation{Microsystems Division, ITC-IRST, via Sommarive 18,
Trento 38050, Italy}

\author{S. Cabrini, and E. Di Fabrizio}
\affiliation{INFM-TASC Laboratory, LILIT Beamline, S.S.14 Km
163.5, 34012, Trieste, Italy}

\begin{abstract}
We present a numerical analysis and preliminary experimental
results on one-dimensional Fabry-Perot micro-cavities in
Si$_3$N$_4$ waveguides. The Fabry-Perot micro-cavities are formed
by two distributed Bragg reflectors separated by a straight
portion of waveguide. The Bragg reflectors are composed by a few
air slits produced within the Si$_3$N$_4$ waveguides. In order to
increase the quality factor of the micro-cavities, we have
minimized, with a multiparametric optimization tool, the insertion
loss of the reflectors by varying the length of their first
periods (those facing the cavity). To explain the simulation
results the coupling of the fundamental waveguide mode with
radiative modes in the Fabry-Perot micro-cavities is needed. This
effect is described as a recycling of radiative modes in the
waveguide. To support the modelling, preliminary experimental
results of micro-cavities in Si$_3$N$_4$ waveguides realized with
Focused Ion Beam technique are reported.
\end{abstract}

\pacs{42.70.Qs, 42.82.Bq, 42.82.Cr, 78.20.Bh}

 \keywords{Waveguide photonic crystals, Fabry-Perot micro-cavities,
 Engineered mirrors, Recycling of leaky modes}

 \maketitle
\newpage

\section{Introduction}

Silicon-based photonics is the key technology for manipulating,
controlling, and detecting light at sub-micrometer length scales
in silicon \cite{SiliconPhotonics}. Owing to the high index
contrast between silicon based materials and air, these systems
are ideal to study devices associated to photonic band-gap
materials. Among these, electromagnetic resonant cavities which
are able to trap light can be considered as building blocks of
future photonic circuits \cite{VahalaNature}. In these systems,
the confinement of the photons within a finite volume is assured
by a periodic refractive index modulation of the surrounding
medium. The best choice is a three-dimensional refractive index
modulation that assures a full confinement of photons in the
cavities. Nevertheless, there are lot of difficulties in
fabricating three-dimensional periodic structures operating at
infrared wavelengths. As a consequence, it seems to be favorable
to explore new devices with two or one dimensional (1D) refractive
index modulation for which the fabrication technology is well
established; light confinement in the other dimensions can be
realized by total internal reflection within optical waveguides.
Photonic crystals based on such an approach exhibit a
quasi-photonic band-gap due to the lack of three dimensional
confinement. An implementation of 1D photonic crystals are
photonic crystal slab waveguides, where a high-index core layer is
sandwiched between lower index claddings. The 1D refractive index
modulation is achieved by producing air trenches across the
channel waveguide. A 1D photonic crystal with a defect is also
described as a Fabry-Perot micro-cavity. This is formed by two
first order Bragg mirrors separated by a spacer. The Bragg mirrors
are constituted by a sequence of $\lambda/2$ thick periods, where
each period contains an air slit and a waveguide segment. A photon
mode propagating in the channel waveguide excites many cavity
radiative modes when it is transmitted trough the cavity. This
fact degrades the quality factor ($Q$-factor) of the cavity.
Indeed the key point for high $Q$-factor cavities relies on a fine
tuning of the first periods facing the cavity spacer
\cite{NodaNature,multipole cancellation,LalanneIEEE} to reduce the
excitation of radiative cavity modes and decrease the impedance
mismatch between waveguide and cavity modes. Nevertheless the
physics beyond these effects is not completely understood, and it
is still under debate \cite{SauvanNature}

The aim of this work is to characterize and optimize the
performance of a 1D Fabry-Perot (FP) micro-cavity centered at 1550
nm and formed in a Si$_3$N$_4$ channel waveguide. We start our
characterization optimizing the insertion losses of a Distribute
Bragg Reflector (DBR) via engineering the first periods facing the
waveguide spacer, as proposed in \cite{LalanneIEEE}. Then we use
the optimized mirrors to build a FP micro-cavity and we study the
quality factor ($Q$-factor) of the optimized system as a function
of the resonance order. In the second part of the work we show
that the classical FP model of $Q$-factor is unable to explain the
numerical results. These can be understood by considering the
recycling of radiative modes in the waveguide
\cite{LalanneOptExp}. In the last part of this work, we report
preliminary experimental data on a FP micro-cavity. The FP
micro-cavity has been realized with Focused Ion Beam (FIB)
processing, starting from a single mode Si$_3$N$_4$ channel
waveguide and removing waveguide slices to define the photonic
structure. Our modelling is able to explain the experimental data.
The full theoretical characterization and realization of other
devices are currently in progress to further validate the
numerical predictions.

\section{Engineered mirror for high $Q$-factor device}

\begin{figure} [t!]
\centering
\includegraphics[width=8.5cm]{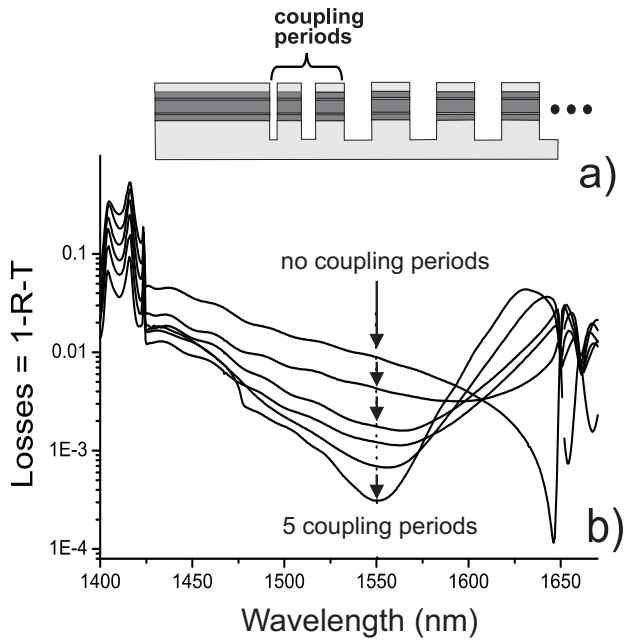}\\
\caption{a) Lateral view of the simulated system composed by the
input waveguide and the mirror where the coupling periods are
emphasized. b) Insertion losses of six optimized mirrors, L=1-R-T,
as a function of the wavelength. The mirrors differ by the number
of coupling periods which face the input waveguide. The arrow is a
guide for the eye to follow the results as the number of coupling
periods increases.} \label{Cavity_tapered}
\end{figure}

Classical FP micro-cavities on waveguides are formed by a straight
waveguide surrounded by two identical DBR. The main limitation of
these systems is the modal mismatch between the fundamental cavity
mode (waveguide mode) and the fundamental Bloch mode of the DBR.
This mismatch causes strong coupling of the resonant guided mode
with radiative modes of the DBR, decreasing the performance of the
micro-cavity \cite{RiboliIEEE}. Thus an optimization of the
coupling between waveguide and DBR is needed. The first step is to
consider the problem related to the reflection of a transverse
electric (TE) mode impinging from a monomodal waveguide onto a DRB
(see Figure~(\ref{Cavity_tapered}a) for a sketch).

As material system we choose Si$_3$N$_4$ waveguides since we are
investigating them as substitute of more expensive silicon on
insulator waveguide \cite{JLTDaldosso}. A monomode Si$_3$N$_4$
waveguide formed by a multilayer core of Si$_3$N$_4$ and SiO$_2$
and a top-bottom cladding of SiO$_2$, will be considered hereafter
(see the experimental section \ref{sec:expresults} for more
details). The waveguide is interfaced to a first order Bragg
mirror with a period of 490~nm which is composed by air slits
(width 100~nm) etched 1$\mu $m down through the bottom cladding.
With these parameters the DBR supports a single guided TE
polarized Bloch mode -- a mode below the light-line. The
characterization of the waveguide-mirror interface is made by
exciting the fundamental waveguide mode and calculating the
reflection $R$ (transmission $T$) coefficient of this mode through
the waveguide-DBR system. The insertion loss spectra, defined as
$L=1-R-T$, are calculated from 1.4~$\mu$m to 1.7~$\mu $m, which
includes the DBR stop band region. All the simulations have been
made with a commercial software \cite{photonD} based on an
eigenmode expansion (EME) method \cite{Sudbo}. The minimization of
the insertion losses is made by means of a multiparametric
optimization tool.

Optimization of the insertion losses is done by inserting coupling
periods at the interface between the waveguide and the DBR. The
physical lengths of each air slit and waveguide segment in the
coupling periods are the free parameters of the multiparametric
optimization. The numerical results of the optimization are shown
in Figure (\ref{Cavity_tapered}b) where the insertion loss spectra
of six optimizations are reported as a function of wavelength. The
six curves correspond to different numbers of coupling periods,
varied from zero (no coupling periods) to five. With increasing
the number of coupling segments a well defined minimum appears at
1.55~$\mu $m; the insertion losses are decreased by two order of
magnitude, from $10^{-2}$ for the bare mirror to $3\times 10^{-4}$
for the five periods engineered mirror. Table~1 reports the
parameters of the coupling periods for the five mirrors. For each
engineered mirror we report the period length $\Lambda$ and the
filling fraction f.f. (the ratio between the length of air slit
and the period) of each set of coupling periods.
\begin{figure}
\centering
\includegraphics[width=6cm]{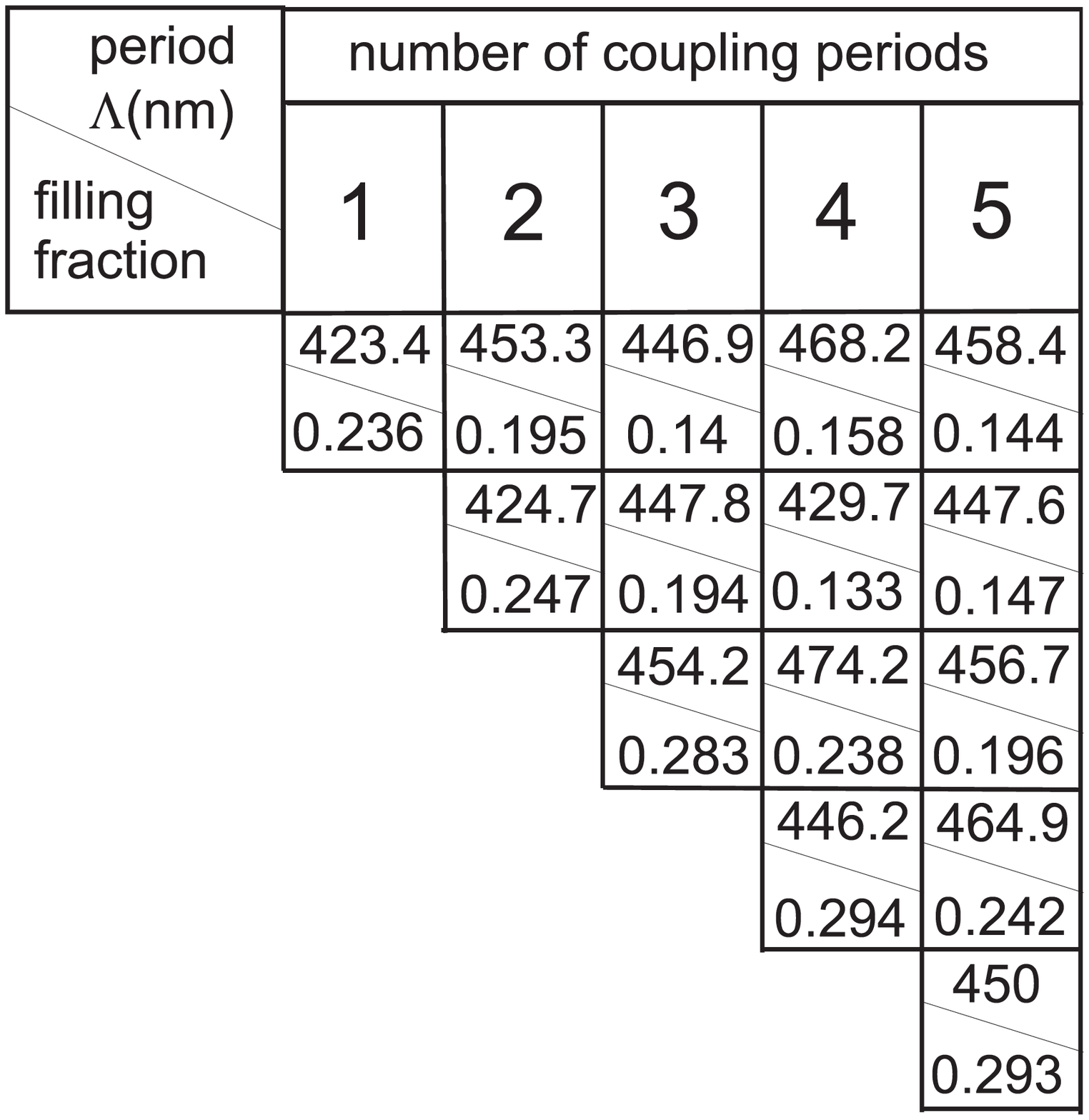}\\
\label{tabella} \vspace{0.2cm}
\parbox{8.5cm}{
Table1: period length $\Lambda$ and filling fraction f.f. of each
set of coupling segments for the five engineered mirror. For the
periodic mirror $\Lambda=490$nm and f.f.=0.204.}
\end{figure}

\begin{figure}
\centering
\includegraphics[width=8.5cm]{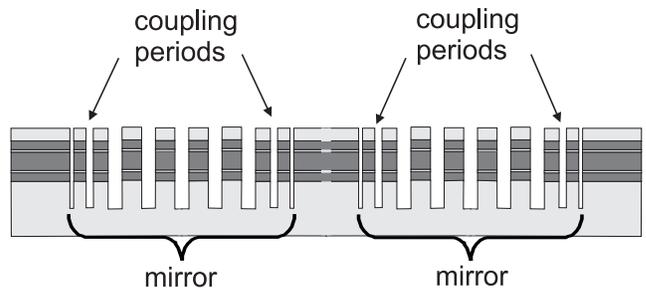}\\
\caption{Schematic lateral view of a micro-cavity with optimized
mirrors. The system is composed by an input-output waveguide, two
engineered mirrors and a spacer. Each engineered mirror has
coupling periods facing the waveguide and the
spacer.}\label{Tapered_Cavity}
\end{figure}

The engineered mirrors have been then used to form optimized
micro-cavities (Figure~\ref{Tapered_Cavity}).
Figure~\ref{Quality_factor} reports their calculated $Q$-factor as
a function of the resonance order $m$ when no (circles), one
(triangles) and two (squares) coupling periods are used. The
micro-cavity intrinsic $Q$-factor was calculated from the
asymptotic value of the full width at half maximum $\Delta\lambda$
of the resonant peak as the number of DBR periods increases while
the number of coupling periods is kept constant; knowing the
micro-cavity resonance wavelength $\lambda_0=$ 1.55$\mu$m,
$Q_{int}=\lambda_0/\Delta\lambda$. The resonance order is defined
as in Eq.~\ref{Qfactor1}. For the smallest defect length the
minimum value of $m$ is about $4$, due the large penetration depth
of the mode defect in the mirror (see also the discussion after
Eq.~\ref{Qfactor1}).

It is seen that the $Q$-factor increases as the number of coupling
periods increases, which was expected, since the insertion losses
of the mirrors $L$ decrease as shown in
Figure~(\ref{Cavity_tapered}). For instance at $m\approx 9$ we
have $Q\approx 3\times 10^3,\;1.1\times 10^4,\; 5.1\times10^4$ for
no, one and two coupling periods, respectively. On the other hand,
it is clearly seen that such an increase is not very regular as
$m$ changes. Such a behavior is discussed in the next section.

\section{Interpretation of the simulation results}

Let us consider a FP cavity, with mirrors characterized by complex
reflectivity and transmissivity coefficients,
$r=|r|e^{i\varphi_{m}}$ and $t=|t|e^{i\varphi_{m}}$, respectively;
the spacer between the two mirrors is characterized by a physical
length $L$ and by an effective refractive index $n_{eff}$. The
expression of the cavity transmission is given by the Airy formula
\begin{equation}
t_{FP}=\frac{t^2e^{i\Phi}}{1-r^2e^{i\Phi}}.
\end{equation}
where the phase accumulated by the electromagnetic wave after half
round-trip in the cavity is
\begin{equation}
 \Phi=\frac{2\pi}{\lambda}n_{eff}L+\varphi_{m}.
 \label{TotalPhase}
\end{equation}
Under the assumption that $1-|r|^2\ll 1$, the quality factor of
the micro-cavity can be expressed as
\begin{eqnarray}
Q=\frac{|r|}{1-|r|^2}\left[\frac{2\pi}{\lambda}n_{g}L
-\lambda\frac{\partial
\varphi_{m}}{\partial\lambda}\right]_{\lambda=\lambda_0}\equiv\frac{|r|}{1-|r|^2}m\pi\label{Qfactor1}
\end{eqnarray}
where $\lambda_0$ is the resonant wavelength,
$n_g=n_{eff}-\lambda(\partial n_{eff}/\partial \lambda)$ is the
group index of cavity mode. In the eq. (\ref{Qfactor1}) we have
also identified the resonance order $m$ with the term in square
brackets divided by $\pi$. The second term in brackets is
proportional to the penetration depth in the mirror.
\begin{figure}
\centering
\includegraphics[width=8.5cm]{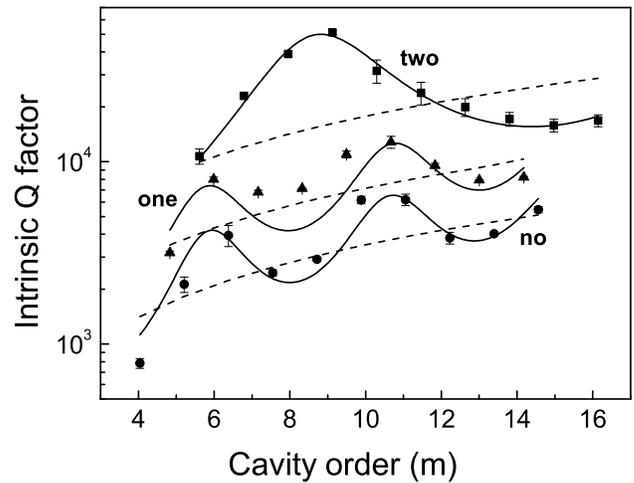}\\
\caption{$Q$-factor of micro-cavities with no (circles), one
(triangles) and two (squares) coupling periods in the engineered
mirrors as a function of the resonance order. Points refer to
numerical results, the dashed lines are the classical FP
prediction and the continuous lines are the predictions from the
FP model with effective reflectivity coefficient for the
fundamental mode.}\label{Quality_factor}
\end{figure}

It is possible to apply this formula to our system, when the
effective refractive index is taken equal to the calculated
effective refractive index of the fundamental mode of the
micro-cavity. The dashed lines in Figure (\ref{Quality_factor})
are the $Q$-factors predicted by eq. (\ref{Qfactor1}) for the
micro-cavities with no, one and two coupling periods in the DBRs,
respectively. Only the order of magnitude is got by this model,
while the oscillations in the $Q$-factors found in the simulations
are not reproduced. It is thus needed to go beyond this simple
model and use more accurate models which predict a non linear
dependence of the $Q$-factor on the mode order. In particular the
theory in Ref. \cite{LalanneOptExp} considers that not only the
fundamental Bloch mode of the micro-cavity is excited by the
fundamental waveguide mode but also radiative cavity modes. What
it is worth of mentioning is that these radiative cavity modes can
be coupled back into the transmitted waveguide mode, a phenomenon
which can be described as a recycling of photons in the waveguide
by the micro-cavity. Hence, not only the fundamental cavity mode
contributes to determine the $Q$-factor of the micro-cavity but
also the radiative ones. The constructive (destructive)
interference among the cavity modes, which eventually increases
(decreases) the $Q$-factor, depends on the cavity parameters
(physical length, modal reflectivity). To simplify the
calculations, the radiative cavity modes are described by a single
leaky cavity mode with two complex parameters, a complex
coefficient for the coupling of the cavity leaky mode with the
fundamental mode and a complex effective index of the leaky mode
itself. Thus, the whole process can be described by an effective
reflectivity coefficient for the waveguide fundamental mode, which
depends on the resonance order $r_{\rm eff}(m)$. The $Q$-factor is
then given by a FP-like expression with the dressed reflectivity
coefficient $r_{\rm eff}(m)$~\cite{LalanneOptExp}
\begin{equation}
Q=\frac{|r_{\rm eff}(m)|}{1-|r_{\rm eff}(m)|^2}m\pi
\label{dressed}.
\end{equation}
The continuous lines in Figure (\ref{Quality_factor}) show the
predictions of this model when the coupling and effective
refractive coefficients are fit to the simulation data. The
agreement in this case is pretty good, suggesting that the
recycling of the leaky mode could play an important role in
determining the performance of such micro-cavities. Even if the
fit parameters have no strict physical meaning, we have checked
that they take values that are compatible with leaky mode
representation: the effective index of leaky mode is lower than
that of the cladding, the real part of the coupling coefficient
satisfies the energy conservation.

\section{Preliminary experimental results}\label{sec:expresults}

To validate the positive role of the coupling periods to increase
the micro-cavity $Q$-factors, we present some preliminary
experimental results for a system realized by FIB technique,
starting from Si$_3$N$_4$ single mode waveguide. As a side-result
we show that FIB technique can be effectively used to form 1D
photonic crystals with good optical quality without the need to go
through the lithography and etching steps used in other
processing.

\begin{figure}
\centering
\includegraphics[width=8.5cm]{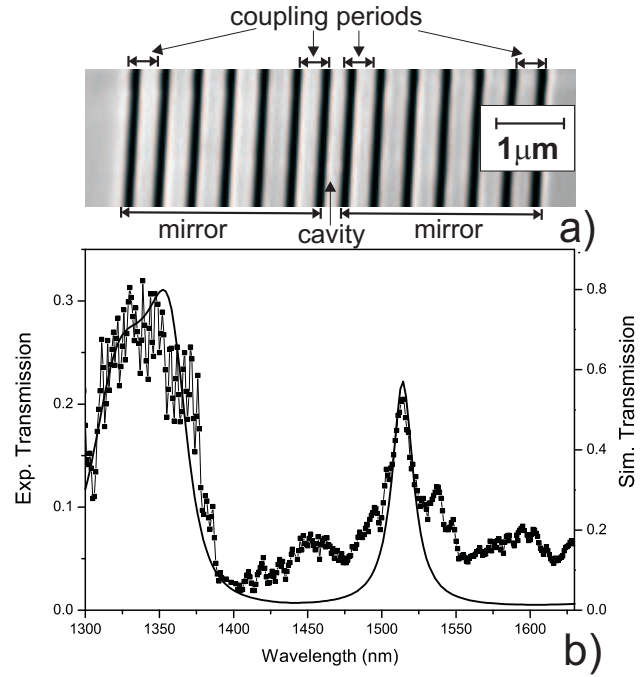}\\
\caption{a) top view SEM image of the cavity with one segment
engineered mirror. b) relative transmittance of transverse
electric (TE) polarized light of the cavity shown in a). The
points shows the experimental data and the continuous line
represent the numerical simulation. The nominal/simulation
parameters are $\Lambda=$435nm/450nm and f.f.=0.23/0.244 for the
coupling period, $\Lambda=$490nm/510nm and f.f.=0.204/0.215 for
the periodic mirror, and 190nm/172nm for the cavity defect.}
\label{Experimental_bis}
\end{figure}

Slab multilayer waveguides were fabricated by low-pressure
chemical vapor deposition (LPCVD) on a 2.5 $\mu$m thick SiO$_2$
and it consists of the following sequence of Si$_3$N$_4$ and
SiO$_2$: 100nm Si$_3$N$_4$, 50nm SiO$_2$, 200nm Si$_3$N$_4$, 50nm
SiO$_2$ and 100nm Si$_3$N$_4$~\cite{Melchiorri_APL}. This results
in a total core layer thickness of about 500 nm. The core was
capped with a 500-nm-thick cladding SiO$_2$ layer. Lithography and
etching defined channel waveguide geometries, whose nominal widths
ranged from 1 to 10 $\mu$m. The 1D-PhC structures were defined on
the channel by using a 30 KeV Ga+ Focused Ion Beam
(FIB)\cite{FIB,Cabrini}. In our experiment, we used the LEO-ZEISS
1540XB CrossBeam$^{\circledR}$, comprising a high resolution FIB
column to mill directly in combination with a high precision
scanning electron microscope (SEM) for precise positioning and
inspection of the fabricated nanostructures in real time. Using an
ion current of about 100 pA and controlling the FIB by a pattern
generator (RAITH ELPHY), the pattern is directly written on the
sample surface; using a total ion dose of 400 mA/cm$^2$, a depth
of about 1.3 $\mu m$ for each structure was obtained.
Figure~(\ref{Experimental_bis}) a shows a top view scanning
electron microscopy (SEM) image of a Fabry-Perot samples with one
coupling period.

The optical properties of integrated optical microcavities have
been characterized by coupling in light from a tunable laser
(1300-1630nm, 2mW) through a single mode polarization maintaining
tapered fiber, mounted on a nano-positioning system. Two linear
polarizers and a half-wave plate are used to control the
polarization of the input signal. The collection system was
provided by a near field microscope objective matched to a
variable zoom mounted on a high-performance InGaAs infrared camera
controlled by the LBA-500 Spiricon beam analyzer software. A prism
beam splitter allows to direct the transmitted signal to a
calibrated photodiode (Ge detector) to perform intensity
measurements. Normalization of the transmission intensity is done
with respect to a nearby reference waveguide without the 1D
photonic crystals.

Figure~(\ref{Experimental_bis}b) shows the relative transmittance
of transverse electric (TE) polarized light for the engineered FP
microcavity of Figure~(\ref{Experimental_bis}a). The measured
spectrum shows a well defined band-edge at 1380 nm and the
resonant peak at 1515 nm. The first FP oscillation (corresponding
to the interference of Bloch-modes  of the 1D system) can be
observed between 1300 and 1380 nm. The measured full width at half
maximum (FWHM) of the resonance peak is 15nm which corresponds to
a quality factor $Q=105$.

The spectrum is compared to the numerical calculation (line in
Fig.~\ref{Experimental_bis}b) and a reasonable good agreement is
found for the spectral position of the FP stop-band, the resonance
position and width, while the baseline in the photonic band-gap
region is not reproduced due to diffraction losses and light
leakages not taken into account in the calculations. The
parameters (length of air slits and waveguide segments) used to
reproduce the spectrum are slightly different from the nominal one
by about 5\%-10\%. The $Q$-factor calculated with classical FP
model is larger by a factor two than the measured one. This can be
a good indication that anti-recycling of the leaky modes into the
fundamental mode occurs in this sample. Clearly, other effects can
affect the measured low $Q$ values which are mostly related to the
quality of the 1D photonic structures, however photon recycling
effects are surely playing a role and only a more systematic study
of these engineered structures could weigh the relative importance
of the photon recycling effect. Such a study is under way.

\section{Conclusions}

In this work we have presented preliminary testing of the effect
of photon recycling which occurs in a 1D photonic crystals when
guided and radiative modes interfere. Simulations and first
measurements show that this effect could not be underestimated in
optimizing the $Q$-factors of micro-cavities based on 1D photonic
crystals.

We acknowledge the financial support by MIUR through FIRB
(RBNE01P4JF and RBNE012N3X) and COFIN (2004023725) projects and by
PAT through PROFILL project.


\begin{thebibliography}{99}

\bibitem{SiliconPhotonics}
L. Pavesi and D. Lockwood (editors), Silicon Photonics, Topics in
Applied Physics vol. 94 (Springer-Verlag, Berlin, 2004).

\bibitem{VahalaNature}
K.J. Vahala, Nature \textbf{424}, 839-846 (2003).

\bibitem{NodaNature}
B.S. Song, S. Noda, T. Asano and Y. Akahane, Nature Materials
\textbf{4}, 207 (2005).

\bibitem{multipole cancellation}
Steven~G. Johnson, S. Fan, A. Mekis, and J.D. Joannopoulos, Appl.
Phys. Lett. \textbf{78}, 3388 (2001).

\bibitem{LalanneIEEE}
P. Lalanne and J.P. Hugonin, IEEE J. Quant. Electr.\textbf{39},
1430 (2003).

\bibitem{SauvanNature}
C. Sauvan, P. Lalanne, J.P. Hugonin, Nature \textbf{429}, 6988
(2004).


\bibitem{LalanneOptExp}
Ph. Lalanne, M. Mias and J.P. Hugonin, Opt. Exp. \textbf{12}, 458
(2004).

\bibitem{RiboliIEEE}
F. Riboli, N. Daldosso, G. Pucker, A. Lui and L. Pavesi,
IEEE-Journal of Quantum Electronics \textbf{41}, 1197 (2005).

\bibitem{JLTDaldosso}
N. Daldosso, M. Melchiorri, F. Riboli, M. Girardini, G. Pucker, M.
Crivellari, P. Bellutti, A. Lui, L. Pavesi, IEEE Journal of
Lightwave Technology \textbf{22}, 1734 (2004).

\bibitem{photonD}
Photon Design Sofware: FimmWave and FimmProp.

\bibitem{Sudbo}
A.S. Sudbo, IEEE Phot. Technol. Lett. \textbf{5}, 342 (1993).

\bibitem{Melchiorri_APL}
M. Melchiorri, N. Daldosso, F. Sbrana, L. Pavesi, G. Pucker, C.
Kompocholis, P. Bellutti, and A. Lui, Appl. Phys. Lett.
\textbf{86}, 121111 (2005).

\bibitem{FIB}
K.A. Valiev, The Physics of Sub-micron Lithography (Plenum Press,
1992).

\bibitem{Cabrini}
S. Cabrini, A. Carpentiero, R. Kumar, L. Businaro, P. Candeloro,
M. Prasciolu, A. Gosparini, L.C. Andreani, M. De Vittorio, T.
Stomeo, and E. Di Fabrizio, Microel. Eng. 78-79, 11 (2005).

\end{thebibliography}
\end{document}